# Enhanced lithium depletion in Sun-like stars with orbiting planets.

Garik Israelian[1, 2], Elisa Delgado Mena[1, 2], Nuno Santos[3, 4], Sergio Sousa[3, 1], Michel Mayor[4], Stephane Udry[4], Carolina Domínguez Cerdeña[1, 2], Rafael Rebolo[1, 2, 5] & Sofia Randich[6]

[1]*Instituto de Astrofísica de Canarias, Via Láctea s/n, E-38200 La Laguna, Tenerife, Spain.*
[2]*Departamento de Astrofísica, Universidad de La Laguna, E-38205 La Laguna, Tenerife, Spain*
[3]*Centro de Astrofisica, Universidade de Porto, Rua das Estrelas, 4150-762 Porto, Portugal.*
[4]*Observatoire de Genève, Université de Genève, 51 ch des Maillettes, CH-1290 Versoix, Swtzerland.*
[5]*Consejo Superior de Investigaciones Científicas, 28006, Madrid, Spain.*
[6]*Instituto Nazionale di Astrofisica, Osservatorio di Arcetri, Largo Fermi 5, I59125 Firenze, Italy.*

**The surface abundance of lithium on the Sun is 140 times less than protosolar[1], yet the temperature at the base of the surface convective zone is not hot enough to burn Li[2,3]. A large range of Li abundances in solar type stars of the same age, mass and metallicity is observed[4,5], but theoretically difficult to understand[3,6,7]. An earlier suggestion[8,9,10] that Li is more depleted in stars with planets was weakened by the lack of a proper comparison sample of stars without detected planets. Here we report Li abundances for an unbiased sample of solar-analogue stars with and without detected planets. We find that the planet-bearing stars have less than 1 per cent of the primordial Li abundance, while**

**about 50 per cent of the solar analogues without detected planets have on average 10 times more Li. The presence of planets may increase the amount of mixing and deepen the convective zone to such an extent that the Li can be burned.**

We obtained Li abundances from high resolution, high signal-to-noise (S/N) spectra for a sample of 451 stars in the HARPS high precision (better than 1 m/s) radial velocity exoplanet survey[11] spanning the effective temperature range between 4900 and 6500 K. These are unevolved, slowly rotating non-active stars from a CORALIE catalogue[11]. These stars have been monitored with high precision spectroscopic observations for years in order to detect planetary systems. Of these 451 stars, 70 are reported to host planets and the rest, which we will designate as a comparison sample, (we often call them "single" stars) have no detected planets so far. If there are planets around these "single" stars, their masses and orbital parameters will be different from those already known. We use this comparison sample to show that the reason for this extra Li depletion is not related to high metallicity (characteristic of planet-host stars) or to old ages.

Our abundance analysis, which followed standard prescriptions for stellar models, spectral synthesis code and stellar parameter determination[12], confirm the peculiar behaviour of Li in the effective temperature range 5600–5900 K for the 30 planet-bearing stars with respect to the 103 stars without planets in the comparison sample. To put this on a more solid statistical foundation these two samples in the $T_\mathrm{eff}$ = 5600–5900 K window were extended by adding 16 and 13 planet-host and comparison sample stars respectively, for which we have obtained new Li abundances from high quality spectroscopic observations using the same spectral synthesis tools. We found that the immense majority of planet-host stars have severely depleted lithium whereas in the comparison sample a large fraction has only partially inhibited depletion. At higher and lower temperatures planet-host stars do not appear to show any peculiar behaviour in their Li abundance. The explanation of lithium survival at $T_\mathrm{eff} \gtrsim 5850$ K is that the convective layers of stars more massive than the Sun are shallow and too remote to reach the Li burning layers. However, lower mass stars with $T_{\rm eff} \lesssim 5700$ K have deeper convective layers that transport surface material to high temperature regions in the interior where Li can be destroyed more efficiently.

The Li abundance of some 20% of stars with exoplanets in the temperature range 5600–5900 K is log $N$(Li) ≥ 1.5 (in standard notation, log $N$(Li) = log $[n({\rm Li})/n({\rm H})]$ + 12, where $n$ is the number density of atoms), while for the 116 comparison stars the Li abundance shows a rather high dispersion with some 43% of the stars displaying Li abundances log $N$(Li) ≥ 1.5. This result becomes more obvious in solar analogue stars where some 50% of 60 "single" stars in the narrow window of $T_{\rm Sun}$± 80 K ($T_{\rm Sun}$=5777 K) appear with log $N$(Li) ≥ 1.5 while only two planet hosts out of 24 have log $N$(Li) ≥ 1.5 (Fig 1). We performed different two-sample statistical tests using ASURV[13] (version 1.2). All tests consistently confirm (at the 3σ level) that the planet-host and single star populations are not drawn from the same parent population. We note that subgiants were not included in this study because they undergo dramatic changes in their internal structure that alters the surface abundance of Li. The Li over-depletion in planet-bearing main sequence stars is a generic feature over the $T_{\rm eff}$-restricted range $T_{\rm Sun}$±80 K and is independent of $T_{\rm eff}$ (or mass). These stars have very similar masses and similar surface convective zone depth, therefore there should be additional reasons for the over-depletion of lithium. We now discuss the impact of age and metallicity on the Li abundance of solar analogue stars.

The lithium abundance of solar-type stars is expected to decrease progressively with age[14,15]. It is in principle possible that solar analogue planet-host stars are on average older than the comparison sample and have depleted more lithium. If that were the case, we should also expect a correlation between lithium and stellar age indicators. Chromospheric activity is a reliable age indicator for solar-type stars from young ages to about[15,16] 1 Gyr, or perhaps even to the age of the Sun[17]. Abundances of Li versus chromospheric activity indices[17], $R_{HK}$, for the solar analogue stars with and without detected planets are shown in Fig. 2a. The comparison of the $R_{HK}$ values for the stars in our sample and for stars in the 625 Myr old[15] Hyades cluster[18] indicates a much older age for our stars. We find no correlation between Li and the activity index in both samples (Fig 2a). This suggests that age is not the main parameter governing Li depletion in our targets. It is known[19] that chromospheric activity correlates with stellar rotation (*v*sin*i*). If the planet hosts were older than the comparison sample, their rotational velocities would be smaller than in the comparison sample. This is not observed either (Fig 2b), adding support to our previous conclusion.

Most of the planet-host stars discovered to date are metal rich[20]. The metallicity excess could result from either the accretion of planets/planetesimals on to the star or the protostellar molecular cloud. This metallicity excess is also present in the solar analogue planet-bearing stars (see Fig. 2c). Can high metallicity be responsible for enhanced Li depletion in these stars? The increase of metal opacities in solar-type stars is responsible for the transition between radiative and convective energy transport. The main contributors to the total opacity at the base of the convective zone are oxygen and iron[21]. Our data (Fig 2c) show that the fraction of single stars with log Li > 1.5 is 50% at both [Fe/H] < 0 and [Fe/H] > 0. This suggests that the Li depletion mechanism does not depend on the metallicity in the range −0.5 < [Fe/H] < +0.5. We have investigated the dependence of log *N*(Li) on [O/Fe] for planet-host stars, using oxygen data from the literature[22], and again found no correlation. Comparison with field stars then leads to the conclusion that neither age nor metallicity is responsible for the excess Li depletion. This is reinforced by observations of Li in solar-type stars in old clusters, which indeed show a wide dispersion of Li abundances with values ranging from log *N*(Li) = 2.5 down to 1.0 and lower[5,23]. This is the case for M67 (age 3.5–4.8 Gyr and [Fe/H] = 0.06)[23] and NGC 6253 (age 3 Gyr and [Fe/H] = 0.35)[24,25], as is clearly seen in Fig 2d. These two clusters offer a homogeneous sample of solar analogues in terms of age and metallicity. Both high and low Li abundance solar analogues are present in these two clusters. The high Li abundance in a large fraction of old metal rich stars in NGC 6253 and M67 leads us to conclude that high

45metallicity and/or age may not be the main cause for the systematic low Li abundances in solar-analogue planet-host stars. Our observations do not suggest that Li is unaffected by metallicity and/or age. They only imply that these parameters are not important enough in order to make the enhanced Li depletion that we observe in solar-analogues with exoplanets.

We propose that the low Li abundance of planet-host solar-analogue stars is directly associated with the presence of planets. The presence of a planetary system may affect the angular momentum evolution of the star and the surface convective mixing. Planet migration will probably increase the angular momentum of the star. Various theoretical studies[3,6,7] show how magnetic braking scales with rotational velocity leading to turbulent diffusion mixing and enhanced lithium depletion. If that were the case we would expect severely Li-depleted stars to host planets with shorter orbital periods. There is no indication for such a correlation in the data, but we also note that in most cases we can only impose upper limits on the Li abundance, so that such correlations with orbital parameters could still be masked in the current data.

Alternatively, a long-lasting star–disc interaction during the pre-main sequence may cause planet-host stars to be slow rotators and develop a high degree of differential rotation between the radiative core and the convective envelope, also leading to enhanced lithium depletion[26]. Revealing the relationships between protoplanetary discs and stellar structure in the early phases of the evolution of solar-type stars is a challenge for evolutionary models and simulations. It is possible that the enhanced Li depletion already takes place in the pre-main sequence stage of planet-host stars. Exoplanet searches in very young stars will be crucial to elucidate this. Asteroseismological observations of solar twins with and without known planets may reveal peculiarities in the inner structure of planet-host stars that could be the key to ascertaining the impact of planetary systems on the structure and angular momentum history of these stars.

It is known that solar-type stars with high metallicity have a high probability of hosting planets. Those solar analogues with low Li content (which is extremely easy to detect with simple spectroscopy) have an even higher probability of hosting exoplanets. Understanding the long-lasting mystery of the low Li abundance in the Sun appears to require proper modelling of the impact of planetary systems on the early evolution of solar analogue stars.

**Acknowledgements** This research has been supported by The Spanish Ministry of Science and Innovation (MICINN). N.C.S. and S.G.S. acknowledge the support from the Fundacao para a Ciencia e a Tecnologia, Portugal, through the programme Ciencia 2007.

**Author Contributions** All authors participated in data collection, analysis, interpretation and commented on the manuscript. G. I. led the project and wrote the paper.

The authors declare that they have no competing financial interests.

**Author Information** Correspondence and requests for materials should be addressed to G. I. (e-mail: gil@iac.es).

Figure 1.

**Lithium abundance against effective temperature in solar-analogue stars with and without detected planets**.

The planet hosts and "single" stars are red filled and empty circles, respectively**.** The red circle with the black point at its center indicates the Sun. The minimum detectable Li abundance varies among the stars used in this study because their spectra have different signal-to-noise ratios. The straight line log $N$(Li)=1.5 matches the upper envelope of the lower limits corresponding to a minimum S/N = 200 in a typical solar twin. We employ this line as a cut-off for selecting Li-depleted stars in our sample. Note that the two planet host stars with the highest Li abundance also have nearly the highest effective temperatures and therefore the thinner convective zones, which help to preserve this element. Apart from these stars, log $N$(Li)=1.5 is the highest value found in a planet-host star. The mean statistical errors (1σ) for the log $N$(Li) and $T_{eff}$ averaged over all stars are 0.06 dex and 30 K, respectively[12]**.** Errors in log $N$(Li) (bottom right corner) include uncertainties in $T_{eff}$ and equivalent width measurement.

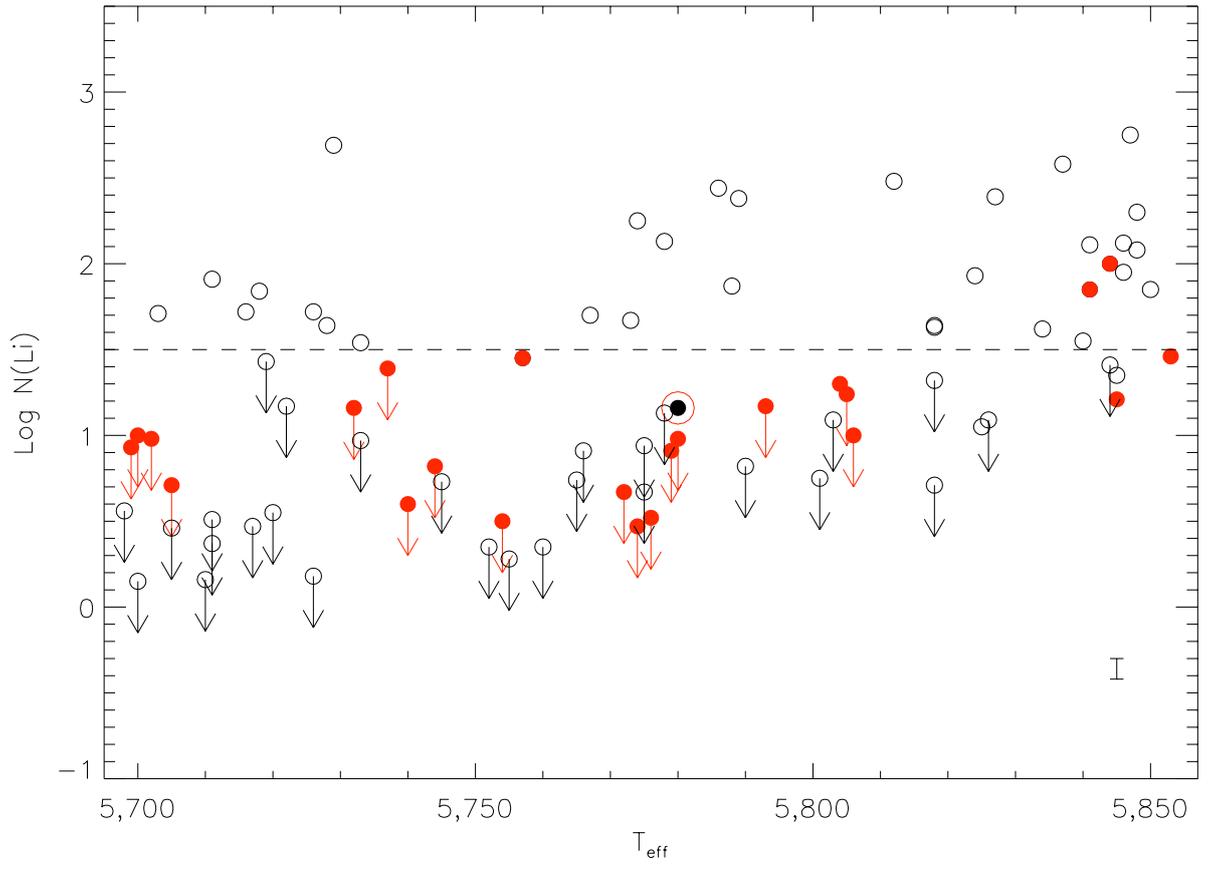

Figure 2.

**Lithium and physical properties of planet-host solar-analogue and comparison stars.**

Chromospheric activity indices $R_{HK}$ were taken from the literature[17, 27, 28] while rotational velocities of the comparison sample stars and many planet hosts were measured from CORALIE and HARPS spectra using a cross correlation function[29]. Typical 1σ uncertainties for log $R_{HK}$ and $v\sin i$ (panels (a) and (b), bottom right corner) are 0.1 dex and 0.3 km sec$^{-1}$, respectively[17,29]. Rotational velocities of several planet-hosts were taken from the literature[29,30]. The metallicities were measured[12] with a 1σ precision of $0.05$ dex (bottom right corner in panel (c)). In the panel (c) we plot Li abundances versus effective temperature in planet-hosts (red filled circles), and stars of the open clusters M67 (blue triangles) and NGC 6253 (open squares). The data for M67 were taken from the literature[15]. Li abundances in NGC 6253 have been derived from VLT/Giraffe spectra using standard methods (Randich *et al.*, in preparation). Typical 1σ error bars (panel (c), bottom right corner) are 0.15 dex and 100 K for log *N*(Li) and $T_{eff}$, respectively.

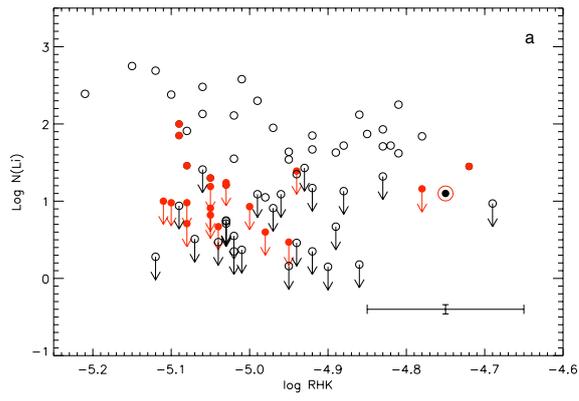
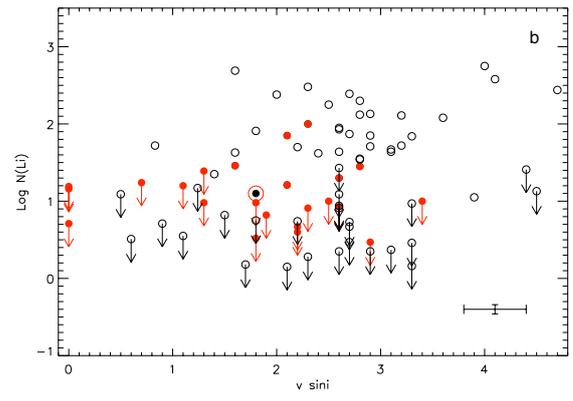
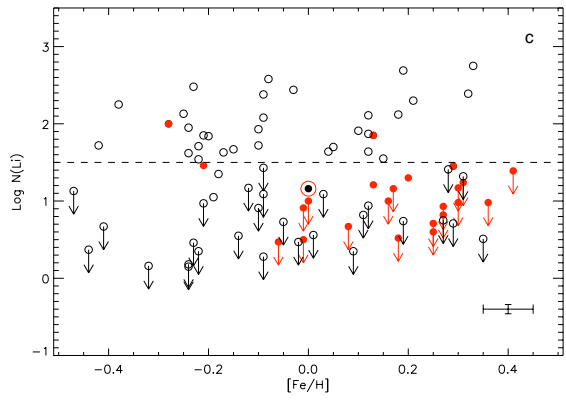
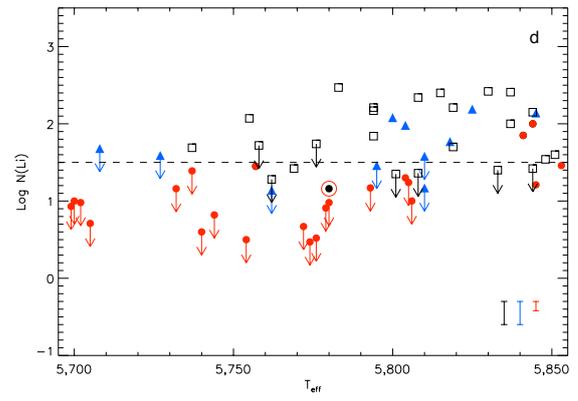